\documentclass[a4paper]{jpconf}
\usepackage{graphicx}
\usepackage{epsfig}
\begin{document}
\def\b{\bar}
\def\d{\partial}
\def\D{\Delta}
\def\cD{{\cal D}}
\def\cK{{\cal K}}
\def\f{\varphi}
\def\g{\gamma}
\def\G{\Gamma}
\def\l{\lambda}
\def\L{\Lambda}
\def\M{{\Cal M}}
\def\m{\mu}
\def\n{\nu}
\def\p{\psi}
\def\q{\b q}
\def\r{\rho}
\def\t{\tau}
\def\x{\phi}
\def\X{\~\xi}
\def\~{\widetilde}
\def\h{\eta}
\def\bZ{\bar Z}
\def\cY{\bar Y}
\def\bY3{\bar Y_{,3}}
\def\Y3{Y_{,3}}
\def\z{\zeta}
\def\Z{{\b\zeta}}
\def\Y{{\bar Y}}
\def\cZ{{\bar Z}}
\def\`{\dot}
\def\be{\begin{equation}}
\def\ee{\end{equation}}
\def\bea{\begin{eqnarray}}
\def\eea{\end{eqnarray}}
\def\half{\frac{1}{2}}
\def\fn{\footnote}
\def\bh{black hole \ }
\def\cL{{\cal L}}
\def\cH{{\cal H}}
\def\cF{{\cal F}}
\def\cP{{\cal P}}
\def\cM{{\cal M}}
\def\ik{ik}
\def\mn{{\mu\nu}}
\def\a{\alpha}

\title{Gravitational strings beyond quantum theory: Electron as a closed heterotic string.}

\author{Alexander Burinskii}

\address{Lab. of Theor. Phys. , NSI, Russian Academy of
Sciences, B. Tulskaya 52  Moscow 115191 Russia}

\ead{bur@ibrae.ac.ru}

\begin{abstract}
The observable parameters of the electron indicate unambiguously
that its gravitational background should be the Kerr-Newman
solution without horizons. This background is not flat and has a
non-trivial topology created by the Kerr singular ring. This ring
was identified with a closed gravitational string. We discuss the
relation of this string to the closed heterotic string of the low
energy string theory and show that traveling waves along the KN
string give rise to the Dirac theory of electron. Gravitational
strings form a bridge between gravity and quantum theory,
indicating a new way to consistent Quantum Gravity. We explain the
pointlike experimental exhibition of the electron and argue that
the predicted closed string may be observed by the novel
experimental method of the "nonforward" Compton scattering.
\end{abstract}

\section{Introduction}
 Modern physics is based on Quantum theory
and Gravity. The both theories are confirmed experimentally with
great precision. Nevertheless, they are conflicting and cannot be
unified in a whole theory. One of the principal contradictions
between Quantum theory and Gravity is the question on the shape
and size of an electron. Quantum theory states that electron is
pointlike and structureless \cite{FWil,LSuss}, and it  is
experimentally supported by the high energy scattering. In the
same time, gravity based on the Kerr-Newman (KN) solution
indicates unambiguously that electron should forms a closed string
of the Compton size. Contrary to the Schwarzschild solution, which
displays the `range'  of a gravitational field (radius of the
horizon $r_g = 2m $) proportional to the mass of the source \fn{We
use the units $\hbar=G= c =1$}, the KN geometry of the rotating
bodies presents a new dimensional parameter $a=J/m ,$ which grows
with angular momentum $J$ and has the reverse mass-dependence. As
a result, the zone of gravitational interaction is determined by
parameter $ a ,$ which increases for the large angular momentum
and small masses, and therefore, it turns out to be essential for
elementary particles. The reason of that, it a specific structure
of the KN gravitational field, which concentrates near the Kerr
singular ring,  forming a closed gravitational waveguide -- a type
of the closed gravitational string \cite{IvBur}.

In 1968 Carter obtained that the KN solution for the charged and
 rotating black holes has $g=2$ as that of the Dirac electron,
 \cite{Car,DKS},
which allowed one to consider KN solution a consistent with
gravity electron model,
\cite{IvBur,Car,DKS,Isr,BurGeonIII,Bur0,Lop,BurSen,BurAxi,BurTwi,BurKN,
BurBag,Dym,TN,BurSol,BurQTS7}. Mass of the electron in the
considered units is $m\approx 10^{-22},$ while $ \ a=J/m \approx
10^{22} .$ Therefore, $a>>m ,$ and the black hole horizons
disappear, opening a twosheeted spacetime with a nontrivial
topological defect in the form of the naked Kerr singular ring.
The Kerr ring takes the Compton radius, corresponding to the size
of a "dressed" electron in QED and to the limit of localization of
the electron in the Dirac theory.

There appear two questions: 1) How does the KN gravity know about
one of the principal parameters of Quantum theory? and 2) Why does
Quantum theory works successfully on the flat spacetime, ignoring
the stringlike peculiarity of the background gravitational field?
A small and slowly varying gravitational field could be ignored,
however, the stringlike KN singularity forms a branch line of the
spacetime and creates a twosheeted topology, ignorance of which
cannot be justified. A simple answer to these questions is to
assume that there is a general underlying theory providing the
consistency of quantum theory and gravity.
 In this paper we suggest an approach which allows one to resolve
 this puzzle from a rather unexpected suggestion, that the
underlying theory is the Einstein-Maxwell gravity as a fundamental
part of the theory of superstrings. In this case, quantum theory
should follow from the theory of superstring, and we make here
first steps in this direction. Starting from description of the
structure of the KN spacetime in sec.II, we show
  in sec.III the relationships of the Kerr singular ring to a heterotic
 string of the low energy string theory, and then, in sec. III we
 show `emergence' of the Dirac equation and the corresponding wave
 functions from the physical model of the lightlike traveling waves,
 propagating along the KN circular ring. In sec. IV we describe a
 model for regularization of the KN solution by the Higgs fields,
 which allows as to regularize the KN background, retaining its
 asymptotical KN form. It justifies the use of flat background
 in the Dirac theory and QED, answering the second above-mentioned
 question.

So far as  gravity predicts the existence of the closed Kerr
string on the boundary of the Compton area, such a string, is
really exists, should be experimentally observable, and there
appears the question while it was not obtained earlier by the high
energy scattering. In Conclusion we give an explanation to this
fact and argue that the KN string may apparently be detected by
the novel experimental method based on the theory Generalized
Parton Distributions (GPD) \cite{Rad,Ji} which represents a new
regime for probing the transverse shape of the particles by the
"nonforward Compton scattering" \cite{Hoyer}.

\section{The Kerr-Newman background} The Kerr-Newman solution in
the Kerr-Schild form has the metric
 \be g_\mn=\eta _\mn + 2 H k_\m k_\n , \label{ksm}\ee
 where \be H=\frac {mr -
e^2/2}{r^2 + a^2 \cos ^2 \theta}, \label{H} \ee and $\eta _\mn $
is metric of auxiliary Minkowski space in the Cartesian
coordinates $(t, x, y, z) ,$ and the Kerr coordinates $r$ and
$\theta$ are the spheroidal oblate coordinates, which are related
to the Cartesian coordinates as follows
\bea \nonumber x+iy  &=& (r + ia) e^{i\phi} \sin \theta  \\
z &=&r\cos\theta . \label{oblate}\eea
 The function $H$
is singular at $r=0, \ \cos \theta=0 ,$ corresponding to the Kerr
singular ring. The KN electromagnetic potential has the form \be
\alpha^\m_{KN} = Re \frac e {r+ia \cos \theta} k^\m\label{ksGA} .
\ee The KN potential as well as the KN metric are aligned with the
null direction $k^\m ,$ which forms the Kerr principal null
congruence (PNC) determined by the differential form \cite{DKS}
 \be k_\m dx^\m=dt +\frac z r dz + \frac {r (xdx+ydy)}{r^2 +a^2} - \frac {a
(xdy-ydx)}{r^2 +a^2} . \label{km} \ee The potential and metric are
singular at the Kerr ring, which forms a branch line of the Kerr
spacetime in two sheets, corresponding to $r>0$ and $r<0 $ in the
Kerr oblate coordinate system. Vector field $k^\m$ forms Principal
Null Congruence (PNC) of KN space, which is determined by the Kerr
theorem in twistor terms, \cite{BurTwi,BurKN}. The Kerr PNC is
smoothly propagated via the Kerr disk $r=0$ from the `negative'
sheet ($r<0$) of spacetime to the `positive' one ($r>0 $) (see
Fig.1), and therefore, it covers the KN space twice: $k^{\m (+)}$
for $r>0$ and $k^{\m (-)}$ for $r<0 , $ leading to different
metrics and different electromagnetic field on the `positive' and
`negative' sheets \cite{BurA}. Therefore, the Kerr ring creates  a
twosheeted background topology.

\begin{figure}[ht]
\centerline{\epsfig{figure=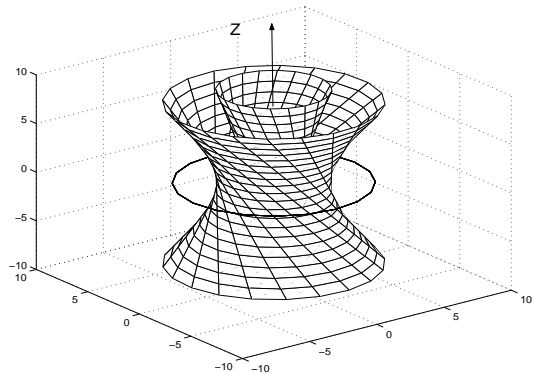,height=5cm,width=7cm}}
\caption{Vortex of the Kerr congruence. Twistor null lines are
focused on the Kerr singular ring, forming a circular
gravitational waveguide, or string with lightlike excitations.}
\end{figure}

This twosheetedness forms a principal puzzle of the Kerr geometry
over a period of four decades. In in 1967 Keres \cite{Keres} and
then Israel \cite{Isr} truncated negative KN sheet, $r<0 ,$,
replacing it by the {\it rotating disklike source} at $r=0 ,$
spanned by the Kerr singular ring of the Compton radius
$a=\hbar/2m .$  Then Hamity assumed in \cite{Ham} that the disk is
to be rigidly rotating, which led to a reasonable interpretation
of the matter of the source as an exotic stuff with the zero
energy density and negative pressure. The matter distribution was
singular at the disk boundary, forming an additional closed string
source, and L\'opez suggested in \cite{Lop} to regularize this
source, covering the Kerr singular ring by a disklike ellipsoidal
surface. As a result the KN source was turned into a rotating and
charged oblate bubble with a flat interior, and further, the
bubble source was realized as a regular soliton-like model
\cite{BurSol}, formed by a domain wall interpolating between the
external KN solution and a flat pseudovacuum state inside the
bubble.

Alternatively, from the very beginning, there were considered the
stringlike models of the KN source, which retained Kerr's
twosheeted topology, forming a closed 'Alice' string
\cite{IvBur,BurKN,BurAxi,Isr2}. The Kerr singular ring was
considered as a waveguide for electromagnetic traveling waves
generating the spin and mass of the KN solution in accordance with
the old Wheeler's "geon" model of `mass without mass'
\cite{BurGeonIII,Bur0,BurGeon0,Wheel}.

\section{The Kerr singular ring as a closed string}
Exact {\it non-stationary} solutions for electromagnetic
excitations on the Kerr-Schild background,
\cite{BurAxi,BurA,BurExa}, showed that there are no smooth
harmonic solutions. The typical exact electromagnetic solutions on
the KN background take the form of singular beams propagating
along the rays of PNC, contrary to smooth angular dependence of
the wave solutions used in perturbative approach!

 Position of the horizon for the excited KS  black holes solutions is determined by
function $H $ which has for the exact KS solutions the form,
\cite{DKS}, \be H =\frac {mr - |\psi|^2/2} {r^2+ a^2 \cos^2\theta}
 \ , \label{Hpsi} \ee where $\psi(x)$ is related to the vector potential
of the electromagnetic field
\be \alpha =\alpha _\m dx^\m \\
= -\frac 12 Re \ [(\frac \psi {r+ia \cos \theta}) e^3 + \chi d \Y
],  \ , \label{alpha} \ee where $\quad \chi = 2\int (1+Y\Y)^{-2}
\psi dY , $ and the vector field $\alpha$ satisfies  the alignment
condition \be \alpha _\m k^\m=0 . \ee The equations (\ref{ksm})and
(\ref{Hpsi}) display compliance and elasticity of the horizon with
respect to the electromagnetic field.

The Kerr-Newman solution corresponds to $\psi=q=const.$. However,
any nonconstant holomorphic function $\psi(Y) $ yields also an
exact KS solution, \cite{DKS}. On the other hand, any nonconstant
holomorphic functions on sphere acquire at least one pole. A
single pole at $Y=Y_i$ \be \psi_i(Y) = q_i/(Y-Y_i) \ee produces
the beam in angular directions \be Y_i=e^{i\phi_i} \tan \frac
{\theta_i}{2} \label{Yi} .\ee

The function $\psi(Y)$ acts immediately on the function $H$ which
determines the metric and the position of the horizon.
 The analysis showed, \cite{BurA},
 that electromagnetic beams have very strong back reaction to metric
 and deform topologically the horizon, forming the holes which
allows matter to escape interior (see fig.3).

The exact KS solutions may have arbitrary number of beams in
different angular directions $Y_i=e^{i\phi_i} \tan \frac
{\theta_i}{2}.$ The corresponding function \be \psi (Y) = \sum _i
\frac {q_i} {Y-Y_i}, \label{psiY}\ee leads to the horizon with
many holes.  For the parameters of an electron, the BH horizons
disappear, however the radiating singular beams appear
unavoidable, and in the far zone the beams tend to the known exact
singular pp-wave solutions. The  considered in \cite{Multiks}
multi-center KS solutions showed that the beams are extended up to
the other matter sources positioned  at infinity.

The stationary KS beamlike solutions may be generalized to the
time-dependent wave pulses, \cite{BurAxi}.

 Due to the factor
$\frac 1 {r +ia \cos\theta}$ in the vector potential $\alpha ,$
any electromagnetic excitation of the KN geometry generates the
traveling waves along the Kerr singular ring (at $r=\cos \theta=0
$), and simultaneously, the pole in $Y$ creates an `axial'
singular beam, which is topologycally coupled with the Kerr ring,
see Fig.2. Therefore, the EM excitations on the KN background have
a paired character, creating simultaneously a `circular' traveling
wave and the coupled with it  a propagating outward `axial'
traveling wave.\fn{ Similar spinor solutions were obtained on the
KN background in the frame of the coupled Einstein-Maxwell-Dirac
equations, \cite{BurSuper}. The resulting massless spinor
solutions turn out to be singular at the Kerr ring and the
fermionic wave excitations generate simultaneously the traveling
waves along the Kerr ring and the coupled `axial' singular
pp-waves.}
 The vector
field $k^\m$ is constant along the `axial' beams, and
asymptotically (by $r\to \infty$) the beams tend to the well known
pp-wave (plane fronted) solutions, for which $k_\m$ is a
covariantly constant Killing direction. Adapting the z-axis of the
coordinate system along the asymptotic pp-wave direction, and
using the light-cone Cartesian coordinates $u=(t+z)/\sqrt 2 ,
\quad v=(z - t)/\sqrt 2 $ and $\z=(x+iy)/\sqrt 2 , \quad \Z=(x
-iy)/\sqrt 2,$ one writes the outgoing pp-wave metric in the form
\be ds^2 = 2du dv + 2 d\z d\Z + 2H (v,\z, \Z) dv^2 .
\label{KSPPwave} \ee

\begin{figure}[ht]
\centerline{\epsfig{figure=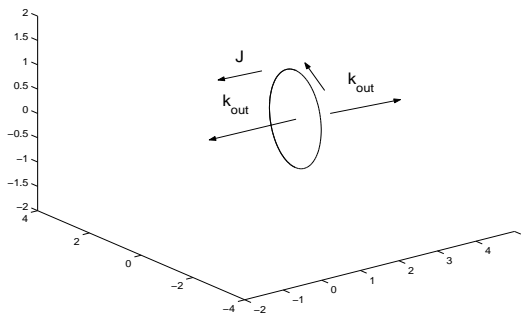,height=5cm,width=7cm}}
\caption{Skeleton of the Kerr geometry \cite{BurAxi} formed by the
topologically coupled  `circular' and `axial' strings.}
\end{figure}

 The pp-waves take important role in superstring theory.
 In the nonperturbative approach based on analogues between the
 strings and solitons, the pp-wave solutions are considered as
 fundamental strings  forming fundamental classical solutions to
 the low-energy string theory
 \cite{WitAxi,HorowSt,DabhGibHarvRR,SenPRL,DablGHarvWald}. The
pp-waves may carry traveling electromagnetic and gravitational
waves which represent propagating modes of the fundamental string
\cite{Garf}. In particular, the generalized pp-waves represent the
singular strings with traveling electromagnetic waves
\cite{BurAxi,Tseyt}.

The string solutions are compactified to four dimensions, and the
singular pp-waves are regarded as the massless fields around a
lightlike singular source of the fundamental heterotic string.
Indeed, metric of the fundamental string solution for traveling
waves in z+ direction takes the form \cite{DablGHarvWald}, \be
ds_{str}^2 = e^{2\phi}(2du dv + 2H (v,\z, \Z) dv^2) + 2 d\z d\Z ,
\label{stringwave} \ee which differs from (\ref{KSPPwave}) only by
factor $e^{2\phi}$ determined by the extra dilaton field $\phi ,$
which deforms the stringy metric in the longitudinal null
direction (note, that the used in string theory `stringy' metric
$ds_{str}^2$ differs from the used in gravity `Einstein' metric
$ds_{E}^2$ by the conformal rescaling
$ds_{str}^2=e^{2\phi}ds_{E}^2$). In particular, for $\phi=0$ the
both metrics are equivalent. The ``solution-generating
transforms'' \cite{SenHeter} allowed Sen to get corresponding
charged string with the lightlike moving current and oscillations,
which was interpreted as a charged (and superconducting) heterotic
string \cite{DablGHarvWald}. There appears also an extra axion
field and the resulting dilaton field turns out to be singular at
the string core.

It is suspected that the singular source of the string will be
smoothed out in the full string theory which should take into
account all orders in $\alpha '.$ Meanwhile, it was shown that the
pp-waves have remarkable property that all the $\alpha '$-
corrections in the string equation of motion are automatically
zero \cite{HorowSt}.

It has been noticed that the field structure of the Kerr singular
ring is lightlike \cite{Bur0}, similarly to the closed fundamental
or the heterotic string \cite{BBS}. The twisted Kerr congruence
(Fig. 1.) represents a ``hedgehog'' defocused by the rotation. The
null lines of the Kerr congruence are focused only in equatorial
plane ($z=\cos \theta=0$). The Eq. (\ref{km}) shows that PNC takes
near the ring the form \be \tilde k =k |_{r=\cos \theta=0}= dt -
(xdy-ydx)/a = dt -a d\phi ,\ee and therefore, the lightlike vector
field $k_\m$ near the ring is tangent to the Kerr string world
sheet, and the Kerr ring is sliding along itself with the speed of
the light. As a consequence, the vector potential (\ref{ksGA}) and
the metric (\ref{ksm}) near the Kerr ring are aligned with the
local direction of the Kerr string reproducing the structure of
the closed pp-string. Therefore, the electromagnetic excitations
of the KN background create the circular traveling waves along the
Kerr ring coupled with the `axial' outgoing lightlike pp-beams.

This similarity of the KN ring with the closed string
 is not incidental, since many solutions to the Einstein-Maxwell theory
turn out to be particular solutions to the low energy string
theory with a zero (or constant) axion and dilaton fields. Indeed,
after compactification the bosonic part of the action for the
four-dimensional low-energy string theory takes the form,
\cite{ShapTriWilc},
 \be S = \int  d^4 x \sqrt {-g} ( R -2 (\d \phi)^2 - e^{-2\phi} F^2
 -  \frac 12 e^{4\phi}(\d a)^2 - a F\tilde F) \label{Seff}, \ee
and contains the usual Einstein term
 \be S_g = \int  d^4 x \sqrt {-g} R \ee completed by the kinetic term for dilaton field
 $ -2 (\d \phi)^2 $ and by the scaled by $e^{-2\phi}$ electromagnetic
 part $ e^{-2\phi} F^2.$ The last two terms are related with axion field $a$ and
 represent its nonlinear coupling with dilaton field \be - \frac 12 e^{4\phi}(\d
 a)^2\ee
 and interaction of the axion with the dual electromagnetic field
\be \tilde F_\mn =
 \epsilon_\mn ^{\lambda \rho} F_{\lambda \rho}.\ee
It follows immediately that {\it any solution of the Einstein
gravity, and in particular the Kerr solution, is to be exact
solution of the effective low energy string theory} with a zero
(or constant) axion and dilaton fields.\fn{Situation turns out to
be more intricate for the Einstein-Maxwell solutions since the
electromagnetic invariant $F^2$  plays the role of the source of
dilaton field, while the term $F\tilde F$ turns out to be the
source of the axion field.} The relationship between the classical
cosmic superconducting strings and the heterotic superstrings was
first mentioned by Witten in \cite{WitAxi}.  The KN string is
closed, charged and has the lightlike current, what are the
characteristic features of  the heterotic strings.\fn{The
characteristic peculiarities of the heterotic strings were
identified by Witten
 with fundamental cosmic strings in \cite{WitAxi}. There was also
mentioned their relation with axion field and  positioning at the
boundary of  a domain wall.}

The stringy analog to the Kerr-Newman solution with nontrivial
axion and dilaton fields was obtained by Sen \cite{SenPRL}, and it
was shown in \cite{BurSen}, that the field around the singular
string in the `axidilatonic' Kerr-Sen solution is very similar to
the field around the heterotic string. The usual KN metric written
in the Kerr-Schild tetrad form is \be ds_E^2 = 2 e^3 e^4 + 2 e^1
e^2 ,\label{gKNtetr} \ee where $e^3$ and $e^4$ are real, and $e^3$
is directed along the Kerr congruence, $e^{3\m} \sim k^\m ,$ while
$e^1$ and $e^2$ are the two complex conjugate null vectors
orthogonal to $ k^\m .$ The metric of the axidilatonic Kerr-Sen
solution has the form (contrary to (\ref{stringwave}) we use here
the usual Einstein metric) \be ds_E^2 = 2 e^3 \tilde e^4 + 2 e^1
e^2 e^{-2(\Phi - \Phi_0)}, \label{axidilmetric} \ee  which shows
that the KN metric is deformed by the dilaton field
$e^{-2(\Phi-\Phi_0)}$ is the
 orthogonal to the Kerr congruence directions.
 Near the Kerr ring, the lightlike direction $e^3$ is tangent
 to the ring, and the Kerr-Sen metric is to be deformed by dilaton in
 the orthogonal to the ring direction.
 The explicit form of the dilaton
 \be e^{-2(\Phi-\Phi_0)}= 1+ \frac{Q^2} {M (r^2 + a^2 \cos ^2
 \theta )} \ee shows that it is singular near the Kerr ring, where $\cos
 \theta =0, \ r \ \to \ 0 ,$ and therefore, metric of the closed
 heterotic string of the Kerr-Sen solution is singular.

This indicates that the Kerr singular ring forms really a
heterotic string  with has to acquire the lightlike traveling
modes.
 The structure of the
Lagrangian (\ref{Seff}) shows that the axion field involves the
dual magnetic field, and therefore, the complex axidilaton
combination $\lambda = a + i e ^{i\phi}$ may generate  the duality
rotation  creating a twist of the electromagnetic traveling waves.
Exact solutions of this type represent especial interest, but so
far are unknown. Note that the complex axidilaton field $\lambda$
was interpreted in F-theory \cite{BBS} as a complex structure of a
torus forming an extra two-dimensional compact space.

The low energy string theory is classical one. Assuming that the
lightlike string forms a core of the electron structure, we have
to obtain a bridge to quantum theory of electron. The lightlike
traveling waves along the KN closed string generate the spin and
mass of the KN particle. Physically, it resembles the original
Wheeler's idea of a "geon'', \cite{Wheel}, in which `mass without
mass' emerges from the system of circulating massless particles.
It hints that the adequate first quantized quantum theory should
be constructed from the initially massless fields, and we show
further, that starting from the massless fields one can derive the
usual Dirac equation for the massive electron.

\section{Traveling waves and `emergence' of the
Dirac equation}

The Wheeler's model of "geon" represents a cloud of the lightlike
particles, which are held by own gravitational field. A
``microgeon'' with the Kerr geometry \cite{Bur0} represents a
degenerate case of a sole `photon' (or other lightlike particle)
circulating on the lightlike orbit of the Compton size. It may be
considered as a corpuscular analogue to traveling waves along the
KN gravitational string.

The observable parameters of the electron indicate unambiguously
the Kerr-Newman background spacetime and the Compton radius of the
corresponding Kerr string. On the other hand, the Compton radius
plays also peculiar role in the Dirac theory, as a limit of
localization of the wave packet. Localization beyond the Compton
zone creates a "zitterbewegung" affecting {"\emph{...such
paradoxes and dilemmas, which cannot be resolved in frame of the
Dirac electron theory...}"} (Bjorken and Drell, \cite{BjoDr}).
Dirac wrote in his Nobel Prize Lecture : \emph{"The variables
$\alpha$ (velocity operators, AB) also give rise to some rather
unexpected phenomena concerning the motion of the electron. .. It
is found that an electron which seems to us to be moving slowly,
must actually have a very high frequency oscillatory motion of
small amplitude superposed on the regular motion which appears to
us. As a result of this oscillatory motion, the velocity of the
electron at any time equals the velocity of light." }

These words by Dirac may be considered as a direct hint regarding
the inner structure of the electron, the core of which should
therefore be represented by a  circulating lightlike particle, and
consequently, its wave function should satisfy the {\it massless}
wave equation, while the mass should appear as an averaged energy
of the circular motion. It corresponds also to generation of the
mass in stringy models from excitations of the traveling modes.
\subsection{Wheeler's microgeon -- mass without mass}
The local 4-momentum of a massless particle circulating around
z-axis has to be lightlike, \be p_x^2 + p_y^2 + p_z^2 = E^2
\label{Ephot} ,\ee  while the effective mass-energy has to be
related with an averaged orbital motion,
 \be <p_x^2> +<p_y^2> = \tilde m^2
\label{mPxy} .\ee The lightlike particle circulating along the
Kerr ring of radius a will generate the angular momentum \be
\tilde J_z=Ea \equiv \tilde ma ,\ee which is in agreement with the
Kerr relation $J_z=ma $ providing consistent source of the mass
$\tilde m=m$ and angular momentum for the KN solution , $\tilde
J_z =J_z.$ The wavelength of the lightlike particle will be \be
\lambda= 2\pi \hbar/E \equiv 2\pi\hbar/m \equiv 2\pi a \hbar /J_z
\label{lambda} .\ee For the half-integer z-projection of the spin,
$J_z=n\hbar/2, \ n=1,2,3..,$ the traveling waves along the KN ring
have the half-integer wavelengths. The ratio \be \frac
 {2 \pi a} \lambda   =  J  / \hbar \ee shows the number of the wave-length
 trapped by the KN ring. For
 $J_z=\hbar /2 ,$ we obtain that traveling mode should have one-half
 of the wavelength corresponding to a two-valued wave function, or
 to be treated on a covering twosheeted manifold.

Averaging (\ref{Ephot}) under the condition (\ref{mPxy}) yields
the relation  \be <p_x^2 + p_y^2 + p_z^2> = \tilde m^2 +p_z^2 =
E^2 \label{mPzE} .\ee The simplest quantum analog of this model
corresponds to the wave equations obtained from the operator
version of these relations: $ \vec p \to \hat {\vec p} = -i\hbar
\nabla , \quad \hat E= i \hbar \d_t .$

\subsection{Massive wave solutions of the scalar massless equation}
We start from the solutions of the scalar massless equation \be
\d_\m \d^\m \phi =0 \label{ddphi} ,\ee which should satisfy the
wave analog of the constraint (\ref{mPxy}), \be (\d_x^2 +
\d_y^2)\phi = - (\tilde m/\hbar)^2 \phi \label{ddxym} . \ee For
the considered orientation of spin, the system of the equations
(\ref{ddphi}) and (\ref{ddxym}) is equivalent to the system \be
(\d_x^2 + \d_y^2)\phi = - (\tilde m/\hbar)^2 \phi = (\d_t ^2 -
\d_z^2)\phi \label{msep} ,\ee which shows explicitly that the
variables may be separated by the ansatz
 \be\phi ={\cal{M}}(x,y)\Phi_0 (z,t)
\label{ans}, \ee where $\Phi_0$ is the wave function of a massive
particle which satisfies  $(\d_t ^2 - \d_z^2)\Phi_0 =- (\tilde
m/\hbar)^2 \Phi_0 $ or equivalently  \be (\d_\m \d^\m - (\tilde
m/\hbar)^2 )\Phi_0 (x,y) =0 \label{ddphi0m} .\ee The corresponding
plane wave solution \be \Phi_0 (z,t) =\exp{\frac i \hbar (z p_z
-Et)} , \label{deBr} \ee has the usual de Broglie periodicity. In
the same time, the l.h.s. of (\ref{msep}) yields the equation \be
(\d_x^2 + \d_y^2){\cal{M}}(x,y) = - (\tilde m/\hbar)^2
{\cal{M}}(x,y) \label{phiInt}, \ee which determines a factor of
``internal vortex structure'' ${\cal{M}}(x,y) .$ The corresponding
cylindrical solutions may be obtained in the polar coordinates
$\rho, \theta .$ From the relation $x+iy = \rho e^{i\theta}$ we
have \be \rho = \sqrt {(x+iy)(x-iy)}, \quad e^{i\theta} = \sqrt
{\frac {x+iy} {x-iy}} \label{polar} ,\ee and the solutions of
(\ref{phiInt}) are expressed via the Hankel functions of order $\n
,$
 ${\cal{H}}_\n ( \frac {\tilde m } \hbar \rho) $
 \be
{\cal{M}}_\n={\cal{H}}_\n (\frac {\tilde m} \hbar \rho)
 \exp \{i\n \theta \} \label{MHan}, \ee which are eigenfunctions
 of the operator
$\hat J_z = \frac \hbar i \d_\theta $ with eigenvalues $J_z=
\n\hbar .$ For electron we put $J_z= \pm \hbar/2, \quad \n=\pm 1/2
,$ and the corresponding functions \be {\cal{M}}_{\pm 1/2}=
\rho^{-1/2}\exp \{ i (\frac {\tilde m} \hbar \rho \pm \frac 12
\theta )\} \label{M12} \ee turn out to be anti-periodic in
$\theta.$ One sees that the wave functions acquire singular ray
along the vortex axis $z ,$ which forms the branch line of
spacetime. The half-integer spin appears here from topological
reason.

We introduce operators \be \hat m_\pm = -i\hbar \d_\pm \equiv
-i\hbar (\d_x \pm \d_y) \label{mpm} \ee  and,  using the relations
\be \frac 1 \rho (\d_x +i\d_y) \rho = \frac 1 {x-iy}, \quad  (\d_x
+i\d_y) \theta = \frac i {x-iy} ,\label{dpm} \ee obtain that the
equation (\ref{phiInt}) may be split into the pair \bea
\hat m_+{\cal{M}}_{-1/2}(x,y) = \tilde m {\cal{M}}_{1/2}(x,y) \nonumber \\
\hat m_-{\cal{M}}_{1/2}(x,y) = \tilde m {\cal{M}}_{-1/2}(x,y) ,
\label{splitM}\eea which is similar to splitting of the Dirac
equation.

The wave function $\phi ,$ solution of the massless equation
(\ref{ddphi}) is factorized into the plane wave solution $\Psi_0
(z,t)$ of the massive Klein-Gordon equation, and the extra
singular factor ${\cal{M}}_\n ,$ which turns the wave function
into a singular string playing the role of a carrier of the de
Broglie plane wave $\Psi_0 (z,t).$ The resulting solution
reproduces the old de Broglie wave-pilot conjecture.\fn{Note that
 structure of this solution is absolutely different from the well
 known Bohm model of the double solution.}

 The obtained solutions may be
 generalized in diverse directions. First of all,
 there may be obtained the corresponding wave solutions, which are
 eigenfunctions of the operator of the total angular momentum $\bf J^2$ and
 simultaneously of the spin projection operator $\bf J_z .$  The treatment
 of the spin in terms of the Pauli-Lubanski pseudovector
 $W^\m=\frac 12 \epsilon_{\mn \rho \sigma} J^{\n\rho} p^\sigma$ and
 the Casimir invariant $W^2 = W_\m W^\m $ allows one to consider the
 above solutions in a Lorentz covariant form for arbitrarily
 positioned and oriented wave functions.
 And finally, the corresponding  solutions of the massless spinor equation
may be obtained.
\subsection{Massive wave solutions of the Dirac massless equation}
Solutions of the Dirac massless equation
 \be
\gamma^\m \d_\m \psi \label{Dirm0}=0 \ee may be constructed from
the obtained above scalar wave functions ${\cal{M}}_\n.$ It is
well known, that if a function $\phi(x)$ satisfies the equation $
\d_\m \d^\m \phi =0 ,$ the corresponding spinor solution \[ \psi =
\left(\begin{array}{c}
 A \\
 B \\
 C\\
D
\end{array} \right) \phi(x) \] with arbitrary coefficients $A,B,C,D$ will
satisfy the massless Dirac equation (\ref{Dirm0}).
 In
particular, using the relations {\ref{mpm}), one can show that the
known two basic plane-wave solutions of the Dirac equation\fn{We
use here notations of the book \cite{BjoDr}.} \be \gamma^\m \d_\m
\psi^r_D = m \psi^r_D \label{Dirm} \ \ee corresponding to positive
energy $E>0 ,$ \be \psi_D^r (x) = w^r(p) \exp \{- \frac i \hbar
p_\m x^\m \} , \ee where $p_\m= (E,0,0,p_z) ,$

 \be w^1 = \left(\begin{array}{c}
 1 \\
 0 \\
 \frac {p_z} {E+m}\\
0
\end{array} \right), \quad w^2 = \left(\begin{array}{c}
 0 \\
 1 \\
 0\\
- \frac {p_z} {E+m}
\end{array} \right) ,\ee
being modified by the singular function ${\cal{M}}_\n (x,y) ,$
yield the  wave functions \bea \psi^1 (x) &=& {\cal{M}}_{-1/2}
(x,y) w^1(p) \exp \{- \frac i \hbar p_\m x^\m \} \nonumber,
\\ \psi^2 (x) &=& {\cal{M}}_{1/2} (x,y) w^2(p) \exp \{- \frac i
\hbar p_\m x^\m \}, \label{MD0}\eea satisfying the massless Dirac
equation (\ref{Dirm0})\fn{One sees that the new spinor components
acquire explicit dependence $e^{i\pm \frac 12 \theta}$ from angle
$\theta .$ It corresponds to the spinor representation of the
Lorentz group for the case when the axis of rotation is given
explicitly \cite{BLP} (Ch.3, sec.23 and App.a).}. Like the
massless scalar solutions, the wave functions of the massless
Dirac equation form a singular string, modulated by the Dirac
plane wave solution.

The performed separation of the variables allowed us to obtain the
ordinary Dirac theory with the massive Dirac equation and the
ordinary Dirac plane wave solutions from the solutions of the
 corresponding massless underlying theory, which regards zitterbewegung as a
 corpuscular analog of the  traveling waves. Note also that the considered
 separation of the variables is exact analog of the usual
 Kaluza-Klein compactification. However, there is a
 principal difference that the  ``compactification''
 is performed in 4d spacetime, in which the Kerr ring plays the role of
 an extra dimension, performing the model of
 a ``compactification without compactification".

\subsection{Impact of the KN gravity and topology}

Like the usual Dirac theory, the above treatment was performed  on
the flat background, ignoring Kerr-Newman gravity. There appears
important question, how these solutions could be deformed by the
KN gravitational field. The result may be elucidated by the known
method of complex shift (see \cite{BurKN,BurSuper,BurNst}). This
method, initiated by Appel in 1887 , was used by Newman at al. to
obtain KN solution from the Kerr one, and it was shown in
\cite{Bur0} that the KN electromagnetic field and the wave
excitations of the KN solutions may be obtained by the complex
shift from the corresponding Coulomb solutions and spherical
harmonics. Application of the complex shift to the empty Minkowski
space displays strong deformation of its causal structure. In
particular, spinor structure of the light cone, which describes a
spherically symmetric hedgehog of the light-like directions, is
deformed into a twosheeted structure of the Kerr Congruence which
should be described by the Kerr theorem in twistor terms. It
follows that the two deforming effects -- influence of the
gravitational field and impact of the twosheeted topology -- are
indeed separated, and the KN background creates the Kerr string
and twosheeted topology even in the limit of the zero
gravitational field (zero mass of the source). On the other hand,
the role of the KN gravitational field turns out  to be very
essential too. The known exact solutions on the Kerr-Schild
background show that the necessary condition for the consistency
with gravity is alignment of the all fields to the Kerr Principal
Null Congruence (PNC) \cite{DKS,BurA,BurExa}. It puts very hard
restrictions on the structure of the`solutions. In fact, the usual
plane waves turn out to be inconsistent with the Kerr-Schild
gravity and should be replaced by the pp-wave analogs, i.e. by
fundamental strings, and, as it was described above, the wave
excitations take the form of the coupled excitations of the axial
and circular strings. Even the very weak  Kerr-Schild gravity has
very strong impact on the structure of the wave solutions leading
to their strong localization near the fundamental strings.  As a
result, in the KN gravity appears new dimensional parameter -- the
Compton wavelength $a =J/m$, and the impact of the very weak
gravitational field turns out to be essential on the Compton
scale, which is very far from the claimed usually Planck scale.

\section{Regularization: Electron as  gravitating KN soliton}
`Emergence' of the Dirac equation from gravitational model of the
traveling waves circulating along the closed Kerr string indicates
a principally new point of view that gravitation and superstring
theory may form some more fundamental theory lying beyond the
Dirac theory of electron. On the level of the low energy string
theory, the Dirac field losses its meaning of the wave function
  and should be considered as a source of the charge and currents for the Maxwell
  equations. The fundamental heterotic strings are charged. The charges
  are  distributed freely along closed strings and localized singularly
  at a core of the strings.
  The mechanism of such localization is
going beyond the  low energy string theory and belongs to an
(unknown so far) effective theory of a more high level.
 As we have
seen, metric of the KN and Kerr-Sen solution turns out to be
singular too, and the raised in introduction question: ``why does
Quantum theory works successfully on the flat spacetime, ignoring
the stringlike peculiarity of the background gravitational
field?'' has left unanswered. To justify consistency of Quantum
theory with the experimentally observable parameters of the
electron, there should also be performed a ``regularization of the
KN or the Kerr-Sen solution'' which should retain the asymptotic
form of the KN solution. Regularization of the BH solutions
represents a very old problem which is close related with the more
general problem of the regularization of the black hole
singularity \cite{Dym0} and with the old problem of the regular
source of the KN solution \cite{BurBag,BEHM,GG}.

 It is often mentioned that the heterotic strings are to be
superconducting \cite{WitAxi,SenHeter,DablGHarvWald}, which
assumes the presence of the mechanism of broken gauge symmetry
realized by the Higgs field. A hint on the important role of the
Higgs field is also coming from the standard model. However, the
Higgs field is absent in the Dirac theory of electron, in QED and
also in the usual version of the low energy string theory.
Meanwhile, the $U(1) \times \tilde U(1)$ Witten field model for
superconducting cosmic strings \cite{WitStr} suggests that the
Higgs field may effectively be used for regularization of the
singular core of the heterotic strings. A supersymmetric version
of the $U(1) \times \tilde U(1)$ Witten field model was applied
for regularization of the KN solution in the gravitating soliton
model \cite{BurSol}, which was mentioned in sec.II. Detailed
treatment of this model goes out of the frame of this work, and in
the next section we consider it briefly, concentrating on its
\emph{gravitational aspect}  and on the \emph{interaction of the
KN electromagnetic (EM) field with the Higgs field}.

The model of gravitating soliton \cite{BurSol} represents a
regular version of the L\"opez bubble model \cite{Lop}. The KN
source forms a domain wall  separating external KN solution from
the flat interior of the bubble which replaces the former Kerr
singular ring. Therefore, interior of the bubble is flat and  the
Kerr singularity is removed.

\emph{Gravitational aspect} of this model is related with a smooth
metric, which interpolates between the external KN solution and
the internal flat spacetime.
 The bubble boundary is determined from the equation $H=0 ,$
 (\ref{H}), which yields \be r=r_e = e^2/(2m) ,\ee where
$r$ is the Kerr ellipsoidal radial coordinate, (\ref{oblate}). As
a result, the bubble takes the form an oblate disk of the Compton
radius $ a =\hbar/(2m)$ with the thickness $r_e = e^2/(2m) ,$
corresponding to the known 'classical size' of the electron. The
KN electromagnetic field is regularized since the maximal value of
the vector potential  is realized in the equatorial plane
$\cos\theta=0 , $ on the stringy boundary of the bubble
\be|\alpha^{(str)}_\m | \le e/r_e = 2m/e . \ee Note, that radius
of the regularized closed string, being shifted to the boundary of
the bubble, turns out to be slightly increased. This position of
the string confirms the known suggestions that the heterotic
strings have to be formed on the boundary of a domain wall
\cite{WitAxi}.

\emph{Chiral sector.} The domain wall should provide a smooth
phase transition from the external KN solution to a flat false
vacuum state inside the bubble. This phase transition is  formed
by a supersymmetric system of the chiral fields $\Phi^i,$
\cite{BurSol} and by the potential $V(r)$ generated from the
specially adapted superpotential $W ,$ which provides  the smooth
transfer from the external KN `vacuum state', $ V_{ext}=0 ,$ to a
flat internal `pseudovacuum' state, $ V_{int}=0 .$  Note, that the
considered chiral field model is a supersymmetric version of the
suggested by Witten $U(1)\times \tilde U(1)$ chiral field model
for superconducting strings \cite{WitStr}, see details in
\cite{BurSol,BurBag}. Therefore, the assumptions that the
heterotic strings are to be superconducting \cite{WitAxi} are also
confirmed in this model.

\emph{Interaction of the Higgs field with the KN EM field} is
controlled by the system of equations \be \d^\n \d_\n \alpha_\m
=I_\m = e |\Phi|^2 (\chi,_\m + e \alpha_\m) \label{Main},\ee which
are equivalent to the  equations used by Nielsen and Olesen for
the stringlike vortex in a superconductor, \cite{NO}. The Higgs
field $\Phi=|\Phi| \exp \{i \chi \} $  forms a superconducting
condensate inside the bubble, which regularizes the KN EM field by
mechanism of broken symmetry. Inside the bubble, the KN EM
potential $\alpha^\m$ is eaten by the Higgs field and the current
$ I_\m$ should vanish there, $ I_\m =0.$
 It gives  \be \d^\n \d_\n \alpha_\m =0,\quad
\chi,_\m + e \alpha_\m =0 \label{In} , \ee which shows that
gradient of the Higgs phase  $\chi,_\m$  compensates the gauge
field $\alpha_\m .$ The field strength and currents are expelled
to the string-like boundary of the bubble, and there appear the
following two conditions: \be \omega=\chi ,_0 = - e
\alpha^{(str)}_0, \quad \chi , _\phi = -e \alpha^{(str)}_\phi
\label{Q0} .\ee We have seen earlier, that the spacelike component
$\mathbf{\alpha}^{(str)}$ of the KN  potential $\alpha^{(str)}_\m
= (\alpha^{(str)}_0, \mathbf{\alpha}^{(str)})$ is tangent to the
Kerr string. As a consequence, it forms a regular flow in
$\phi$-direction, $ \alpha^{(str)}_{\phi} ,$ and creates a closed
Wilson line along the Kerr string. As a result of (\ref{Q0}), the
KN gravitating soliton exhibits two important peculiarities:
\begin{itemize}

\item the quantum Wilson loop $ e \oint  \alpha^{(str)}_\phi
d\phi=-4\pi ma \label{WL} $  leads to quantization of the total
angular momentum of the soliton, $J=ma=n/2, \ n=1,2,3,...$,

\item the Higgs field inside the bubble turns out to be
oscillating with the frequency $\omega=2m ,$ and therefore, it
forms a coherent vacuum state, typical  for the `oscillon' bubble
models.
\end{itemize}
The negative sheet of the metric disappears and metric turns out
to be regularized and practically flat. The model contains the
lightlike heterotic string on the border of the bubble, however,
it is axially symmetric and the traveling waves are absent.
Therefore, some extra excitations are needed to create the
traveling waves.\fn{This point has interesting consequences which
can be interpreted in the terms of quarks, and we expect to
considered it elsewhere.}
 The Compton size of the bubble does not contradict to the
``dressed'' QED electron, however, there is an essential
difference. The dynamics of the virtual particles in QED is
chaotic and can be conventionally separated from the
``bare''electron, while the vacuum state inside the KN soliton
forms a {\it coherent oscillating state} joined with the closed
Kerr string. Therefore, the bubble source of the KN soliton and
the Kerr String represent an {\it integral coherent structure}
which cannot be separated from the ``bare'' particle.

\section{Conclusion.}
We have showed that gravity definitely indicates presence of a
closed string of the Compton radius $a=\hbar/(2m)$ in the electron
background geometry. This string has gravitational origin and is
closely related  with the fundamental heterotic string of the low
energy string theory. Starting from the corpuscular aspect of the
underlying model of the massless traveling modes along the Kerr
string, we showed `emergence' of the usual Dirac equation and the
corresponding `massive' solutions with de Broglie periodicity. The
original Dirac theory is modified in this case, and the wave
functions acquire the singular stringlike carriers, so that the
Dirac plane wave solutions turn out to be propagating along
`axial' singular strings. Contrary to the widespread opinion that
quantum theory predominates over all other theories, the
considered structure indicates opposite: gravity, as a basic part
of the underlying superstring theory, is to be lying beyond
quantum theory. As we have seen, in the KN gravity appears new
dimensional parameter -- the Compton wavelength $a =J/m ,$ and
\emph{the impact of the very weak gravitational field turns out to
be essential on the Compton scale, which is very far from the
claimed usually Planck scale.} The Kerr-Newman solution, together
with interpretation of its source as a closed heterotic string,
forms a bridge between gravity, superstring theory and the Dirac
quantum theory, displaying a new way towards unification of
quantum theory and gravity.

The observable parameters of the electron show unambiguously that
the background of the electron should be determined by the KN
geometry and should contain the Kerr string of the Compton radius.
This radius is very big with respect to the modern scale of the
experimental resolution, and it seems that this string should be
experimentally detected. However, the high-energy scattering
detects  the pointlike structureless electron  down to $10^{-16}
cm $. One of the reasons of this fact was considered earlier in
\cite{BurTwi}, where it was argued that the pointlike character of
the interactions may be caused by the interactions of the KN
particles via the considered above `axial' pp-strings -- carriers
of the wave function. There are also diverse alternative
explanations, for example, related with a known complex
interpretation of the KN solution
\cite{Bur0,BurSen,BurKN,BurSuper}, which shows that the real KN
solution is generated by a complex pointlike source shifted in the
imaginary direction.
 Here we presents a new  explanation of this fact related with
 the lightlike character of the closed heterotic string. Observation
 of the lightlike, relativistically rotating closed string should
 be accompanied by the \emph{ Lorentz contraction} for each line element of
 the closed string, and as a result, to a shrinking of the full image
 of this string to a point in the ultrarelativistic limit.

Indeed, observation of the lightlike objects is very nontrivial
process which depends essentially on the method of the
observation. In particular, it has been shown by Penrose
\cite{Pen} that a \emph{momentary photo-image} of a relativistic
sphere shouldn't display the Lorentz contraction at all. Similar
effect for a meter stick was obtained still earlier by Terrel
\cite{Ter}. It doesn't mean that the Einstein relativity is wrong,
rather, the effect of the Lorentz contraction depends essentially
from the method of observation. We note that the relativistic
scattering differs absolutely from the \emph{momentary
photo-image}, and visibility of the Lorentz contraction by the
relativistic scattering on the closed relativistic heterotic
string is a-priori unclear.

For an intermediate \emph{real} photon, propagating between the
separated points $P_1$ and $P_1 ,$ the usual null condition for
the lightlike interval  $s_{12}^2=0 $ is valid. Such a scattering
 corresponds to an `action-at-a-distance' and should lead
to a pointlike, `contact' interaction which doesn't allow one to
determine the form of a string or its size. An alternative regime
may be related with scattering of a virtual photon, for which  the
strong lightlike condition $s_{12}^2=0 $ is broken. Specifically,
to recognize the stringlike relativistic object of a big size,
there are necessary two especial conditions:

a) the scattering should be {\it deeply virtual}, which means very
large $Q^2 = ­ q_{12}^2 , $ and $ p \cdot q_{12} ,$

b) the momentum transfer  should be  {\it relatively low} to
provide the wavelengths comparable with extension of the object.

Both these conditions are satisfied in  the novel method of
scattering -- the ``non-forward Compton scattering'', or Deeply
Virtual Compton Scattering (DVCS), suggested by Radyushkin and Ji
\cite{Rad,Ji} and broadly discussed last decade. It is projected
now as a new regime of the high energy scattering, which could
realize tomography of  the elementary particles \cite{Hoyer}.

Experimental observation of the predicted closed heterotic string
in the electron structure could have an extraordinary meaning
leading to essential progress in understanding of Quantum theory
and development of Quantum Gravity.

\subsection{Acknowledgments} Author thankful to organizers, and especially
to G. Gr\"ossing, for very kind invitation to attend this
conference and its brilliant organization, and also to many
colleagues for  useful discussions during the series of the summer
2011 conferences, in particular to A. Efremov, O. Teryaev, D.
Gal'tsov, A. Khrennikov, A.Kamenschik, T. Nieuwenhuizen, Yu.
Obukhov, O.Selyugin, K. Stepaniants, and especially to G. `t Hooft
and also to A. Radyushkin for discussions concerning the GPD
application.


\section*{References}
\begin{thebibliography}{9}
\bibitem{FWil} Wilczek F 2008 {\it The Lightness Of Being} (Basic Books)

\bibitem{LSuss} Susskind L 2008 {\it The Black Hole War} (Hachette Book Group
US)
\bibitem{IvBur}  Ivanenko D D and Burinskii A Ya 1975
Gravitational strings in the models of elementary particles {\it
Izv.\ Vuz.\ Fiz.}  {\bf 5} 135


\bibitem{Car} Carter B 1968 Global structure of the
Kerr family of gravitational fields {\it
  Phys.\ Rev.} {\bf 174} 1559


\bibitem{DKS}  Debney G C, Kerr R P and Schild A 1969
  Solutions of the Einstein and Einstein-Maxwell equations {\it
  J.\ Math.\ Phys.}  {\bf 10} 1842

\bibitem{Isr}
Israel W 1970 Source of the Kerr metric {\it  Phys.\ Rev.}  D {\bf
2} 641

\bibitem{BurGeonIII}   Burinskii A Ya 1972 Microgeon with Kerr metric
{\it Abstracts of the III Soviet Gravitational Conference} (
Yerevan, 1972, p.217, in Russian).



\bibitem{Bur0}   Burinskii A Ya 1974 Microgeons with spins {\it
Sov.\ Phys.\ JETP } {\bf 39} 193



\bibitem{Lop} L\'opez C A 1984 An extended model of the electron in general relativity
  {\it Phys.\ Rev.}  D {\bf 30} 313


\bibitem{BurSen} Burinskii A 1995
Some properties of the Kerr solution to low-energy string theory
{\it  Phys. Rev.}  D {\bf 52}, 5826 [arXiv:hep-th/9504139]

\bibitem{BurAxi}
Burinskii A 2004 Axial stringy system of the Kerr spinning
particle {\it Grav. Cosmol.} {\bf 10} 50  [arXiv:hep-th/0403212]


\bibitem{BurTwi} Burinskii A 2004 Twistorial analyticity and three stringy
systems of the Kerr spinning  particle {\it Phys.\ Rev. }  D {\bf
70} 086006 [arXiv:hep-th/0406063]


\bibitem{BurKN}
Burinskii A The Dirac-Kerr-Newman electron 2008 {\it  Grav.\
Cosmol.} {\bf 14} 109
  [arXiv:hep-th/0507109]

\bibitem{BurBag} Burinskii A 2002 Supersymmetric Superconducting Bag as a Core
of Kerr Spinning Particle {\it Grav. Cosmol.} \textbf{8} 261
[arXiv:hep-th/0008129]



\bibitem{Dym} Dymnikova I 2006 Spinning superconducting electrovacuum soliton
 {\it Phys.\ Lett.}  B {\bf 639} 368


\bibitem{TN}  Nieuwenhuizen Th M 2007 The Electron and the Neutrino as Solitons
in Classical Electrodynamics In: {\it Beyond the Quantum} (eds.
Th.M. Nieuwenhuizen et al., World Scientific, Singapure)
pp.332-342


\bibitem{BurSol} Burinskii A 2010 Regularized Kerr-Newman solution as a
gravitating soliton, {\it J. Phys. A: Math. Theor.} {\bf 43}
392001 [arXiv: 1003.2928].



\bibitem{BurQTS7} Burinskii A 2012 Gravity vs.
 Quantum theory:  Is electron really  pointlike? {\it J. Phys.:
Conf. Ser.} (to appear)  [arXiv:1112.0225]



\bibitem{Rad}  Radyushkin A V 1997 Nonforward parton
distributions {\it Phys. Rev.} {\bf D 56} 5524 [hep-ph/9704207]


\bibitem{Ji} Ji X 1977 Gauge-invariant decomposition of nucleon spin
{\it Phys. Rev. Lett.} {\bf 78 }, 610  [hep-ph/9603249].

\bibitem{Hoyer} Hoyer P and Samu Kurki Samu 2010  Transverse shape of the
electron {\it Phys. Rev.} {\bf D 81} 013002 [arXiv:0911.3011].


\bibitem{BurA} Burinskii A 2009 First Award Essay of GRF:
   Instability of black hole horizon with respect to electromagnetic
  excitations  {\it Gen.\ Rel.\ Grav.} {\bf 41} 2281
  [arXiv:0903.3162 [gr-qc]].


\bibitem{Keres} Keres H 1967 To physical interpretation of the
solutions to Einstein equations {\it Zh.Exp. i Teor.Fiz (ZhETP)}
{\bf 52} 768 (in Russian)



\bibitem{Ham} Hamity V 1976 An "interior" of the Kerr metric
{\it Phys. Lett.} A {\bf 56} 77

\bibitem{Isr2} Israel W 1977  Line sources in general relativity, {\it
Phys. Rev.} {\bf D 15} 935


\bibitem{BurGeon0}   Burinskii A Ya 1971 On  the particlelike solutions of
the massles wave equations. {\it Abstracts of  the VIII All-Union
Conference on Elementary Particle Theory} (Uzhgorod, January 1971
pp.96-98 in Russian)



\bibitem{Wheel} Wheeler J A 1962 {\it Geometrodynamics} (Academic
Press, New York)

\bibitem{Multiks}
Burinskii A 2005 Wonderful consequences of the Kerr theorem  {\it
Grav. Cosmology}  {\bf 11} 301  [hep-th/0506006]


\bibitem{GSW} Green M B,  Schwarz J H and Witten E 1987 {\it Superstring
Theory V. I}  (Cambridge University Press)


\bibitem{WitAxi} E. Witten, ``Cosmic Superstrings'',
Phys. Lett. B {\bf 153}, 43 (1985).


\bibitem{HorowSt}  Horowitz G and  Steif A 1990 Spacetime
Singularities in String Theory {\it Phys.\ Rev.\ Lett.}  {\bf 66}
260

\bibitem{DabhGibHarvRR} Dabholkar A, Gibbons G, Harvey J and
Ruiz Ruiz F 1990  Superstrings and Solitons {\it Nucl. Phys.} {\bf
B 340} 33


\bibitem{SenPRL} Sen A 1992 Rotating charged black hole
solution in heterotic string theory {\it Phys.\ Rev.\ Lett.}  {\bf
69} 1006

\bibitem{DablGHarvWald} A. Dabholkar, J. Gauntlett, J. Harvey and
D. Waldram, ``Strings as Solitons \& Black Holes as Strings'',
Nucl.Phys. {\bf B 474}, 85 (1996)  [arXiv:hep-th/9511053].


\bibitem{Garf} Garfinkle D 1992 Black string traveling waves {\it
Phys.\ Rev.} {\bf D  46} 4286

\bibitem{BurSuper} A. Burinskii, ``Super-Kerr-Newman solution to broken N = 2
supergravity'',  Class. Quant. Grav. {\bf 16} 3497 (1999),
[arXiv:hep-th/9903032v2],


\bibitem{Tseyt} Tseytlin A 1995 Exact solutions of closed string
theory {\it Class. Quant. Grav.} {\bf 12} 2365
[arXiv:hep-th/9505052]

\bibitem{SenHeter} Sen A 1992 Macroscopic Charged Heterotic
String {\it Nucl.Phys.} {\bf B 388} 457  [arXiv:hep-th/9206016].

\bibitem{BBS} Backer K, Backer M and Schwarz John H 2007 {\it String Theory and M-Theory: A Modern
Introduction} (Cambridge University Press, ISBN 0-521-86069-5)


\bibitem{ShapTriWilc}  Schapere A, Trivedi S and Wilczek F 1991
 Dual dilaton dion {\it Modern Phys. Letters} A. {\bf 6} 2677



\bibitem{BjoDr}  Bjorken J and  Drell S 1964 Relativistic Quantum
Mechanics (McGraw Hill Book)


\bibitem{BLP} Berestetskii V B, Pitaevskii L P and Lifshitz E M
1971-74 {\it Relativistic quantum theory}  v.2 (Oxford, New York,
Pergamon Press)



\bibitem{BurNst} Burinskii A 2003 Complex Kerr geometry and nonstationary Kerr solutions.
{\it Phys. Rev. D} {\bf 67}, 124024 .


\bibitem{BurExa} Burinskii A 2010 Fluctuating twistor-beam solutions
and holographic pre-quantum Kerr-Schild geometry {\it J. Phys.:
Conf. Ser.} {\bf 222} 012044  [arXiv:1001.0332]


\bibitem{Dym0} Dymnikova I 2006 Vacuum Nonsingular Black Hole
{\it Gen.Rel.Grav.} {\bf 24} 235


\bibitem{BEHM} Burinskii A, Elizalde E,  Hildebrandt S R and
Magli G 2002 {\it Phys. Rev. D} {\bf 65}, 064039
[arXiv:gr-qc/0109085]

\bibitem{GG}
G\"urses M and G\"ursey F 1975 {\it J. Math. Phys.} {\bf 16} 2385



\bibitem{WitStr} Witten E  1985 Superconducting strings  {\it Nucl.
Phys.} {\bf B 249} 557

\bibitem{NO}
Nielsen H B and Olesen P 1973 {\it Nucl. Phys.} {\bf B61} 45


\bibitem{Pen} R. Penrose, "The Apparent shape of a
relativistically moving sphere",  Proc. Cambridge Phil. Soc. {\bf
55} 1371 (1959).

\bibitem{Ter} J. Terrel, ``Invisibility of the Lorentz
Contraction'', Phys. Rev. {\bf 116} 1041 (1959).


\end{thebibliography}
\end{document}